\documentclass{JHEP3}
\usepackage{graphicx}

\newcommand{\be}{\begin{equation}}
\newcommand{\ee}{\end{equation}}
\newcommand{\bea}{\begin{eqnarray}}
\newcommand{\eea}{\end{eqnarray}}
\newcommand{\IR}{\mathbb{R}} 

\def\IZ{\relax\ifmmode\hbox{Z\kern-.4em Z}\else{Z\kern-.4em Z}\fi}
\newcommand{\IS}{{\bf S}}

\newcommand{\non}{\nonumber \\}\begin{Large}\end{Large}

\def\half{{1 \over 2}}

\def\Sch{Schwarzschild}

\def\gm{\gamma}

\def\hlm{{\hat \lambda}}

\def\tgm{{\tilde \gamma}}

\newcommand{\sbsection}[1]{\vspace{.5cm} \noindent {\it #1}}

\def\Tn{{\bf T}^n} \def\T1{{\bf T}^1}
\def\vz{\vec{z}}
 \def\vk{\vec{k}} \def\ve{\vec{e}}
\def\({\left(} \def\){\right)}
\def\[{\left[} \def\]{\right]}
\def\Schw{~Schwarzschild~}
\def\B{{\cal B}} \def\tD{\tilde{D}}

\preprint{{\tt gr-qc/0407058}}

\title{\center{On Black-Brane Instability In an Arbitrary Dimension}}

\author{Barak Kol and
  Evgeny Sorkin
  \\
  Racah Institute of Physics\\
  Hebrew University \\
  Jerusalem 91904,
  Israel\\
  {\tt barak\_kol, sorkin @phys.huji.ac.il}}

\abstract{The black-hole black-string system is known to exhibit
  critical dimensions and therefore it is interesting to vary the
  spacetime dimension $D$, treating it as a parameter of the system.
  We derive the large $D$ asymptotics of the critical, i.e. marginally
  stable, string following an earlier numerical analysis.  For a
  background with an arbitrary compactification manifold we give an
  expression for the critical mass of a corresponding black brane.
  This expression is completely explicit for $\Tn$, the $n$
  dimensional torus of an arbitrary shape. An indication is given that
  by employing a higher dimensional torus, rather than a single
  compact dimension, the total critical dimension above which the
  nature of the black-brane black-hole phase transition changes from
  sudden to smooth could be as low as $D\leq 11$.}

\begin{document}

\section{Introduction}

In the presence of extra compact dimensions, there exist several
phases of black objects depending on the relative size of the object
and the relevant length scales in the compactification manifold. More
precisely, by ``black objects'' we mean massive, non-rotating (static)
solutions of General Relativity with no extra matter\footnote{We
  assume these are non-singular outside their event horizon.}.

So far most research on this problem concentrated on the
background $\IR^{D-1} \times \IS^1$, where $\IS^1$ is a single
compact dimension
and often $D$ has taken the values $5,6$. However, we know of two
instances of a critical dimension in this system, where the
qualitative behavior changes. Therefore, we consider spacetimes of
total dimension $D$ out of which $d<D$ dimensions are extended,
and we vary both $D$ and $d$ as parameters of the system.

The evidence is as follows. In the first case \cite{TopChange}
evidence was given that the behavior of the system near the
conjectured point of merger between the black hole and black string
branches depends on a critical dimension $D^*_{\mbox{merger}}=10$,
such that for $D < D^*_{\mbox{merger}}$ there are local tachyonic
modes around the tip of the cone (conjectured to be the local geometry
close to the thin ``waist'' of the string) which are absent for $D >
D^*_{\mbox{merger}}$. The second case relates to a different point in
the phase diagram -- the Gregory-Laflamme point (GL) where the string
is marginally tachyonic \cite{GL1}. It was recently found
\cite{pertString} that the order of transition has a critical
dimension\footnote{Of course dimensions are integral, and the notation
  means only that the change in the order happens between $D=13$ and
  $D=14$.} $D^*_{GL}=$``13.5'', such that for $D<D^*_{GL}$ the
transition is first order, and otherwise it is second order. Moreover,
there is another example, closely related to the first, of a critical
dimension in a different system -- the Belinskii-Khalatnikov-Lifshitz
(BKL) analysis of the approach to a space-like singularity, where
there is a critical dimension $D^*_{BKL}=10$, such that for
$D>D^*_{BKL}$ the system becomes non-chaotic (see the review
\cite{DHN} and references therein). This last example is related to
our system not only in that it displays a critical dimension, but also
in that the equations are similar to those of the merger transition --
the BKL equations for the metric components as time approaches the
singularity are similar to those for the cone perturbations near its
tip.

A description of relevant works follows. Gregory and Laflamme
\cite{GL1,GL2} discovered the critical mass for a uniform string,
$\mu_{GL}$, and determined it (numerically) for $5 \le D \le 10$.
Actually, as was realized by Reall \cite{Reall}, in order to
determine the critical mass it is enough to know the eigenvalue
for the negative mode of the Euclidean \Schw geometry discovered
by Gross Perry and Yaffe (GPY) \cite{GPY} and evaluated there in
$4d$. The analysis of \cite{GPY} was generalized to $d$ dimensions
by Prestidge \cite{Prestidge} (see also \cite{JGregoryRoss} for a
related discussion of black holes in a box).  Gubser \cite{Gubser}
set a method of perturbation analysis around the 5D GL critical
string whose first step is to determine $\mu_{GL}$ by a somewhat
different way than using \cite{GPY}. That method was perfected by
Wiseman \cite{GubserPerfected,Wiseman1} who also applied it in 6D.
Its generalization to an arbitrary dimension was made in
\cite{pertString}, (see appendix \ref{appendix_GL} for a summary).
In fact, by now there is a considerably growing literature on this
black-hole black-string system. See [12-17]
which study non-uniform strings and [18-26]
which concentrate on black holes in the system.

In this paper we obtain some results on the dimension dependence
of the GL effect and in particular of the critical mass. Recall
that the dimensionless mass of a black string is defined by $\mu
:=G_D M/L^{D-3}$ where $G_D$ is the $D$ dimensional Newton
constant and $L$ is the asymptotic length of the compact circle.
In \cite{pertString} the critical mass of a black string on
$\IR^{D-1} \times \IS^1$ was obtained numerically for a range of
$D$'s, up to $D=50$, and a ``phenomenological'' exponential
behavior $\mu_{GL} \propto \gamma^D$ was observed where $\gamma
\approx .68$. In section \ref{large-dimension} we derive the large
$D$ asymptotics of the critical mass and determine that $\gamma =
\sqrt{e/2 \pi}$ in agreement with that ``phenomenology''. It is
interesting to compare this behavior with that of the ``equal
areas for equal mass'' estimator for the critical
string\footnote{This is defined
  to be the mass for which the areas of the uniform black string and a
  localized, naively undeformed black hole are equal.}, $\mu_S$: both
expressions are exponential in $D$ but with a different base since
for $D\gg 1$, $\mu_S \propto \gamma_S^{~D}$ where
$\gamma_S=e^{-1/2} \approx 0.606$.

Moreover, the behavior of the critical dimensionless mass turns
out to be equivalent to a simple behavior of the critical \Schw
radius: $r_0 \simeq (L/2 \pi) \sqrt{d}$, showing that for large
$d$ the critical string is actually ``fat'' and as we show in
section \ref{indication} most probably it cannot decay into a
black hole.

In order to derive the asymptotics we use the equation for the
negative mode of the $d$ dimensional Euclidean Schwarzschild
geometry which is a single, second order, ordinary differential
equation (ODE). In the large $d$ limit the equation becomes the
radial part of the flat space Laplacian with all the non-trivial
information being carried by the boundary condition close to the
horizon.

In section \ref{torus-section} we proceed to vary the dimension of the
compactification manifold and we express the critical mass of a
uniform brane in this general background in terms of the lowest
eigenvalue of the Laplacian operator on that manifold. In particular,
when the compactification manifold is the flat $n$-dimensional torus
of an arbitrary shape the expression can be made explicit in terms of
the shortest vectors in the reciprocal lattice. We note that there are
interesting cases when several modes turn tachyonic simultaneously.

Finally in section \ref{indication} we raise the question whether it
is possible to see the second order behavior for dimensions smaller
than $14$ using $\Tn$ compactifications. It would be especially
interesting if it happened for dimensions $D \le 11$ where
M-theory/string theory is believed to be a consistent theory of
quantum gravity. As in \cite{pertString} we use as an estimator for
the critical dimension the intersection between $\mu_{GL}(D)$ and
$\mu_S(D)$. We find that the critical dimension can indeed be reduced
and for $3\leq n \leq 6$ it is estimated to be around $D=10$ (see
figure \ref{fig_2ndOrder}).

\section{Large dimension asymptotics for $\mu_{GL}$}
\label{large-dimension}

Consider a $D$ dimensional background with $d<D$ extended dimensions,
and the other $D-d$ being compact. In this section\footnote{This
  assumption will be relaxed in later sections.}  $d=D-1$ and the
compactification manifold is $\IS^1$ of period $L$.

A perturbation theory around the static uniform black string in
$D$ dimensions was recently employed in \cite{pertString} to
construct the non-uniform string branch emerging from the critical
GL point. Among other results it was observed that the critical
mass follows essentially an exponential law as a function of $D$
\be \label{mass_scaling} \mu_{GL} \propto \gamma^D, \ee
where numeric analysis yields $\gm \approx 0.686$, the prefactor
is approximately $0.47$, and our definition of the dimensionless
mass is as usual
\be \label{mu} \mu :=G_D M/L^{D-3} ~, \ee
where $G_D$ is the $D$ dimensional Newton constant\footnote{In this
  paper we work in units such that $G_D=1$.}.

This critical mass was compared with $\mu_S$, the commonly used
``equal areas for equal mass'' estimator.  More precisely it is an
estimator for the point of the first order transition, and it is
only an estimate since in the absence of exact solutions for caged
black holes\footnote{By a ``caged black hole'' we mean a black
hole in a
  compactified spacetime.} one approximates the mass-area relation
with that of a small spherical black hole. One finds
\be \label{mu_0} \mu_S= {1 \over 16 \pi}{\Omega_{D-3}^{~D-3} \over
\Omega_{D-2}^{~D-4} }
  {(D-3)^{(D-3)(D-3)} \over  (D-2)^{(D-2 )(D-4)}}~,
\ee
where $\Omega_{D-1}:=D\,\pi^{D/2}/(D/2)!$ is the area of the unit
$\IS^{D-1}$ sphere.

Performing a series expansion for $D \rightarrow \infty$ we obtain
\be \label{log_mu_0} \log(\mu_S) = -\half  D\left(1+ O\left[
  {\log(D) \over D} \right]\right). \ee
Hence, for  large $D$, $\mu_S$ also exhibits an exponential
scaling, though with a different exponent, $\gamma_S=1/\sqrt{e}
\approx 0.606$. Since $\gamma_S < \gamma$,  $\mu_S$ becomes much
smaller than $\mu_{GL}$ as $D$ increases. Moreover, there is a
dimension, $D \approx 12.5$ for which $\mu_S$ intersects
$\mu_{GL}$. This is an indication that for higher dimensions the
black string becomes unstable before the black hole phase can have
superior entropy. This, in turn, can be regarded as a hint that
above that dimension the unstable GL-string decays to a state
different from the black hole, and it is plausible to expect it to
decay into a slightly non-uniform string. And indeed, exact
calculations in higher orders of perturbation around the GL-point
reveal an existence of a critical dimension ($D^*=13.5$) where the
order of the phase transition between the uniform and the
non-uniform black strings solutions changes from the first to
second order \cite{pertString}.

It is desirable to have an analytic formula for the numerical
constant $\gamma$ in (\ref{mass_scaling}).  Unfortunately, the
equations that were solved in \cite{pertString} in order to obtain
the scaling (\ref{mass_scaling}) are two coupled second order,
rather complicated, ODEs that required a numerical treatment. In
the large $D$ limit these equations simplify, but nevertheless we
did not find an analytic solution. Hence, motivated by Reall's
realization \cite{Reall}, we turned to look at the negative mode
of the Euclidian \Schw geometry, which yields a single ODE which
we were able to solve analytically in the large $D$ limit.

\subsection{The negative mode of the  $d$-dimensional \Sch }

We aim  to identify  the negative modes of perturbed Euclidean
\Sch-Tangherlini metric \cite{Tangherlini} in $d$ (extended)
dimensions. The line element for that background is
\bea \label{SchTang} ds^2 = +f(r) dt^2 &+& f(r)^{-1} dr^2 +r^2
d\Omega_{d-2}, \non f(r)&=&1-\({r_0 \over r}\)^{d-3}~, \eea
$t$ is periodic with period $4\pi r_0 /(d-3)$, and in this section
we work in units where $r_0=1$.

The (gauge invariant) spectrum of the perturbations $\phi_{ab}$ in
the transverse-traceless gauge is obtained by solving the
eigenvalue equations
\be \label{Lich} \triangle_L \phi_{ab} =\lambda \phi_{ab}, \ee
where in full index notation the Euclidean Lichnerowicz operator
is
 ${\triangle_L}_{a b}^{ ~~c d}=-\delta_a^c\, \delta_b^d\, \Box - 2 {R}_{a~b}^{~ c ~ d
 }$.

For this spacetime the negative eigenvalue perturbation is static
and spherically symmetric \cite{GPY,Prestidge}. Let us write the
ansatz for such a perturbation as
\be \label{tt_pert} \phi_{a}^{b} = {\rm diag}
\{\psi(r),\chi(r),\underbrace{\kappa(r),\dots,\kappa(r)}_{d-2 ~\rm
{terms} }\}~. \ee
 The traceless condition sets
\be \kappa(r)=-{1 \over d-2} \left[\chi(r)+\psi(r)\right] \ee
while transversality, ${\phi^{a b}}_{;b}=0$, implies
\be
\label{psi-chi}
\psi(r)= {2 r f\over r f'-2 f} \chi'(r) +{ r f' +2 (d-1) f \over  r
  f'-2 f} \chi(r)
\ee
Hence, the  equations (\ref{Lich}) reduce to a single linear
second order ODE for $\chi(r)$
\bea
\label{chi_eq}
-f\chi''(r)&+& \left[{2\, r^2 \,(f\, f''-f'^2) -r \,(d-2)\,f \,f'  +2\, d\, f^2\over
    r\,(r \,f' -2\, f)}\right] \,\chi'(r) \non
&+&\left[{r^2 f' f'' +r\left[2\,(d-1)\,f  f'' -(d+2) f'^2\right] +4 f f'
    \over  r\,(r \,f' -2 \,f)}\right] \chi(r) = -k^2 \chi(r)
\eea
where for convenience we define $\lambda=-k^2$, such that for the
negative eigenvalues $k$ is real. The 4d version of this equation
was analyzed in \cite{GPY}, while in \cite{Prestidge} (see also
\cite{JGregoryRoss}) the generalization to arbitrary $d$ was
considered.

This equation has three singular points in the region outside the
horizon $r \geq r_0$: at the horizon $r=r_0$ itself, at infinity
and at $r=r_s$ which is the solution to $r f'-2 f=0$.
$r_s$ is given by \be
 r_s^{~d-3}={d-1 \over 2}
 \label{rs-def} \ee
and we note that it is exactly the critical radius for trapping
light, namely, a light ray originating from infinity will fall
into the black hole if (and only if) it crosses $r_s$. Regularity
of $\chi(r)$ at the horizon (a regular singular point) and at
infinity (an irregular singular point) determines the solution. At
the horizon, where $f(r_0)=0$ the characteristic exponents are
$-1,0$ namely $\chi \sim (r-r_0)^\sigma$ with $\sigma =-1,0$, and
so regularity of (\ref{chi_eq}) imposes
\be
\label{bc_rh}
\chi'(r_0)/\chi(r_0)= - \left[d - {k^2\over 2(d-3)}\right].
\ee
At $r=r_s$, on the other hand, the characteristic exponents are
$\sigma =0,3$ and both solutions are finite (though with possible
log's). Looking at a generic solution with $\chi \sim (r-r_s)^0$
we find
\be \label{cond_rs} \chi'(r_s)/\chi(r_s)= -{1\over 2 r f} \left[r
f' +2(d-1)
  f\right]_{r=r_s}=-{d \over r_s}.
\ee
It is important to realize that this relation is {\it not} a
boundary condition, but rather a consequence of the equation
itself.  We observe that since the characteristic exponents are
$0,3$, regularity at $r_s$ can be obtained by extending the
equation (\ref{chi_eq}) to become fourth order (such that the
exponents $\sigma=1,2$ are solutions as well),
 which is what happens in Gubser's gauge (see the
Appendix).

\subsection{Large $d$ limit}
In general, the solution to the equation (\ref{chi_eq}) must be
obtained numerically [7-9].
 However, let
us consider this equation in the limit $d\rightarrow \infty$.  In
this limit $f(r) \rightarrow 1$ and $f'(r), f''(r) \rightarrow 0$
for $r>r_s>1$
 and hence the eigenvalues equation (\ref{chi_eq})
simplifies to
\be \label{chi_eq_bessel} \chi''(r) +{d\over r} \chi'(r) -k^2
\chi(r)=0 ~,\ee
which is the flat space Laplacian (in a $d+2$ dimensional spacetime).
In this approximation $r_s$ is not singular anymore and hence
(\ref{cond_rs}) does not hold automatically. Instead we impose it
as a boundary condition \be
 \label{bc_largeD} \chi'(r_s)/\chi(r_s)=
-{d \over r_s}~. \ee A more rigorous treatment would derive this
effective boundary condition by solving the original equation in
the vicinity of $r=1,\, r_s$ and matching it in the large $d$
limit with the solutions of (\ref{chi_eq_bessel}).

We shall now show that the solution to (\ref{chi_eq_bessel})
subject to the boundary condition (\ref{bc_largeD}) and regularity
at infinity can be found analytically.

Changing variables in (\ref{chi_eq_bessel}) by $\chi(r)=u(r) r^{-\nu}$
with $\nu=(d-1)/2$ we obtain for $u(r)$
\be
\label{bessel}
r^2 u''(r) +r u(r) -( \nu^2 +k^2 r^2) u(r) =0
\ee
which is just a modified Bessel equation whose solution is
\cite{Arfken}
\be u(r)= C_1\, K_\nu(k r) +C_2\, I_\nu(k r) \ee
where $I_\nu$ and $K_\nu$ are the modified Bessel functions of the
first and the second kind respectively \cite{Arfken}.  The
asymptotic boundary condition forbids a growing exponential
solution. Hence $C_2=0$ and the selected solution is $K_\nu$,
whose asymptotic behavior is $K_\nu (z) \sim e^{-z}/\sqrt{2 \pi
z}$.
The boundary condition (\ref{bc_largeD}) dictates
\bea \label{eq_k1} x\, K'_\nu(x) &+& (d-\nu)\, K_\nu(x) =0 \non
 x &:=& k\, r_s ~. \eea
Using the recurrence relations for $K_\nu$ one arrives to a simpler
algebraic equation
\be
\label{eq_k2}
 K_\nu(x) -x\, K_{\nu-1}(x)=0 \ ,
\ee
that determines the eigenvalue $k$ as a function of $\nu$. We note
that in order to arrive at this equation one relies on the
cancellation $d-2 \nu=1$, which is a delicate feature of this process.
The leading behavior of $k=k(d)$ can be gotten by making the
self-consistent assumption $k \ll d$, in which case one may use the
leading behavior $K_\nu(x) \simeq \Gamma(\nu)\, (x/2)^{-\nu}/2$ valid
for $x \ll \nu$.  Substituting into (\ref{eq_k2}) we get
\be \label{approx_k} k={1 \over r_s}\, \sqrt{d-3} \simeq \sqrt{d}
~, ~~~~ d\gg 1 ~, \label{k-d} \ee
confirming the assumption $k \ll d$, and at the same time showing
that the critical string is also surprisingly ``fat'' $r_0/L=k/(2
\pi) \gg 1$ at large $d$.

Eq. (\ref{eq_k2}) may be solved as a series in $1/d$ although it
is probably not justified since it represents only the leading $d$
limit of the original equations where various $1/d$ corrections
were already neglected. Still we record that if one uses more
orders in the Taylor series of $K_\nu(x)$ the solution of
(\ref{eq_k2}) can be expanded as
$ x=\sqrt{d} -{1/\sqrt{d}}+ O({1/d^{3/2}})$ for $ d\gg 1  $.
 This formula yields less than $0.2 \%$ discrepancy with the numerical
 solution of (\ref{eq_k2}) for $d=20$ and the discrepancy is even
 smaller for larger $d$.
 Moreover, this estimate differs by less than $4\%$ from the full
 solution of eq.  (\ref{chi_eq}) for $d=50$ and by $2.5\%$ for
 $d=100$ (see Table \ref{table_kc} in appendix \ref{appendix_GL}.)
This is an additional confirmation of  the relationship between
the GL tachyon and the GPY negative mode.

\subsection{The critical mass}

We can get now some insight into the mass scaling (\ref{mass_scaling})
of the marginally stable black strings. From the definition of $\mu$
(\ref{mu}) with $L=2 \pi/k_{GL}$ we get the critical mass for the
GL-instability
\be \label{mu_vs_k} \mu_{GL}={(d-2) \Omega_{d-2}\over 16 \pi}\,
{G_D \over G_d} \, r_0^{~d-3} \left({k_{GL} \over
    2\pi}\right)^{D-3} = {(d-2) \Omega_{d-2}\over 16 \pi}\, \left({r_0\, k_{GL} \over
    2\pi}\right)^{d-3}  ~,
\ee
(recall that $G_D/G_d$ is the volume of the internal manifold which is
$L$ in this section). Expanding $\log(\mu_{GL})$ in series for $d
\rightarrow \infty$ and keeping the leading terms we get
\be \label{log_mu} \log\left(\mu_{GL}(k_{GL})\right) =
 d\, \log \left(\sqrt{e\over 2 \pi}{ k_{GL} \over \sqrt{d}} \right)
 \left( 1 + O\left[{1\over d}\,
\log\left({k_{GL}\over \sqrt{d}}\right)+ {\log(d)\over d}
\right]\right). \ee

Since according to (\ref{k-d}) $k_{GL}\rightarrow\sqrt{d}$ in this
limit, we obtain from the above expansion
\be \label{get_gamma} \log(\mu_{GL})\simeq d\,
\log\left(\sqrt{e\over 2 \pi}\right) \ee
Hence we learn that for large $D=d +1$ the critical mass should
scale as $\mu \propto \tgm^D$ with
\be
 \label{gamma} \tgm = \sqrt{e\over 2 \pi} \approx 0.658 \ee
 The numerical value $\gamma=0.686$ differs from this $\tgm$ by about
 $4\%$. Note, however, that the numerical estimate was obtained for
 relatively small $D$'s \cite{pertString}, which should account for
 this deviation.

\section{Torus compactification}
\label{torus-section}
In this section we determine the critical mass for GL instability of
black branes on the $D$-dimensional background $\IR^d \times {\bf
  T}^n$, namely $d$ extended dimensions and $n$ compactified on a
torus, and $D=d+n$. The torus is completely general, so its period
vectors may be of any size and with arbitrary angles between each
other.

The relevant black branes are the uniform $n$-branes ${\cal
B}_n$={\bf Schw}$\times \Tn$, namely those which do not depend on
the internal coordinates $\vec{z} \equiv z^i,\, i=1,\dots,n$.
Their metric is given by a sum of the $d$ dimensional \Schw metric
(\ref{SchTang}) and the flat torus metric $ds^2_{\Tn} = dz^i\,
dz^i$.
In addition, the torus satisfies the periodicity boundary
conditions: it is defined by a lattice which may be given by $n$
period vectors $\vec{e}_i$ such that any function $Y(\vec{z})$
must satisfy $Y(\vec{z}+\vec{e})=Y(\vec{z})$ where $\vec{e}$
belongs to the lattice: $\vec{e}=\sum_{i=1}^{n}\, m^i\, \vec{e}_i$
with $m^i$ being arbitrary integers.

The critical mass is defined such that a perturbation mode
$h_{\mu\nu}(r,\vz)$ turns marginally tachyonic \be
 \triangle_L h_{\mu\nu}(r,\vz) =0 ~, \label{marg-tachyon} \ee
where $ \triangle_L$ is the Lichnerowicz operator in the total
space ${\cal B}_n$. Separating the compact variables from the
extended ones, and restricting to perturbations which are scalar
with respect to $\Tn$
 we have \be
 h_{\mu\nu}(r,\vz) = h_{\mu\nu}(r)\, Y(\vz) ~.\ee
 The Lichnerowicz operator in the total space decomposes into \be
 \triangle_L h_{\mu\nu}(r,\vz) =Y(\vz)\, \triangle_L h_{\mu\nu}(r)
 + h_{\mu\nu}(r)\, (-\triangle Y(\vz)) ~, \ee
 where $\triangle$ is the Laplacian operator in the compact space
$\Tn$ and $\triangle_L h_{\mu\nu}(r)$ describes the Lichnerowicz
operator on the $d$ dimensional \Schw geometry. Denoting the
eigenvalues by $\lambda_{\mbox{Schw}},\, \lambda_{\Tn}$ \bea
 \triangle_L h_{\mu\nu}(r) &=&
 \lambda_{\mbox{Schw}}\, h_{\mu\nu}(r) \non
 -\triangle Y(\vz) &=& \lambda_{\Tn}\, Y(\vz) ~,\eea
we have that the total eigenvalue (which should vanish by
\ref{marg-tachyon}) can be written as the sum \be
 0=\lambda_{\mbox{Schw}} + \lambda_{\Tn}~. \label{lambda-sum} \ee

In order to find a zero eigenvalue one of the eigenvalues should
be negative. Actually there is only one such mode -- it is the GPY
mode of the \Schw geometry \cite{GPY}. For this reason it is
enough to consider modes which are tensor on the \Schw geometry
and scalar on the torus.
 So we take
\be
 \lambda_{\mbox{Schw}} = - {k_d^{~2} \over r_0^{~2}} \ee
where $k_d^{~2}$ are the eigenvalues defined in (\ref{chi_eq}),
computed numerically in \cite{pertString} and whose large $d$
asymptotics is (\ref{k-d}). Substituting back into
(\ref{lambda-sum}) we find that {\it the critical $r_0$} below
which the black brane destabilizes is \be
 r_0^{~2} = {k_d^{~2} \over \lambda_{\mbox{min}} } \ee
where $\lambda_{\mbox{min}}$ is the minimal (non-zero) eigenvalue
of the Laplacian on $\Tn$. Actually the argument is more general
and {\it applies to any stable compactifying manifold}. In the
case of $\Tn$ this eigenvalue may be described explicitly
 \be
\label{k_Tn}
 \lambda_{\Tn}  = |\vec{k}|^2 \ee
 where $\vk$ is in the reciprocal lattice, namely $\vk \cdot \vec{e} =2
 \pi \, j,\, j \in \IZ$ for any lattice vector $\vec{e}$.
Finding the minimal eigenvalue of the Laplacian reduces to finding
the shortest vector in the reciprocal lattice of $\Tn$.

\sbsection{The qualitative options}

There are interesting (even if degenerate) cases where there are
several vectors in the reciprocal lattice which lie closest to the
origin, for example, a cubic lattice and a triangular one. Clearly
that would mean that several modes would turn marginally tachyonic
at the same brane mass
 \bea
 &\#& (\mbox{marginal tachyons}) = \non
 = &\#& (\mbox{reciprocal lattice vector
 closest to the origin}) := \# \{ \vk: 
 |\vk|^2=|\vk|^2_{\mbox{min}} \} /2 ~,\eea
where in the last expression $\vk$ is in the reciprocal lattice
and the division by $2$ accounts for the trivial degeneracy of
$\vk$ and $-\vk$.

\sbsection{An example:  ${\bf T}^2$}

\begin{figure}[t!]
\centering \noindent
\includegraphics[width=9cm]{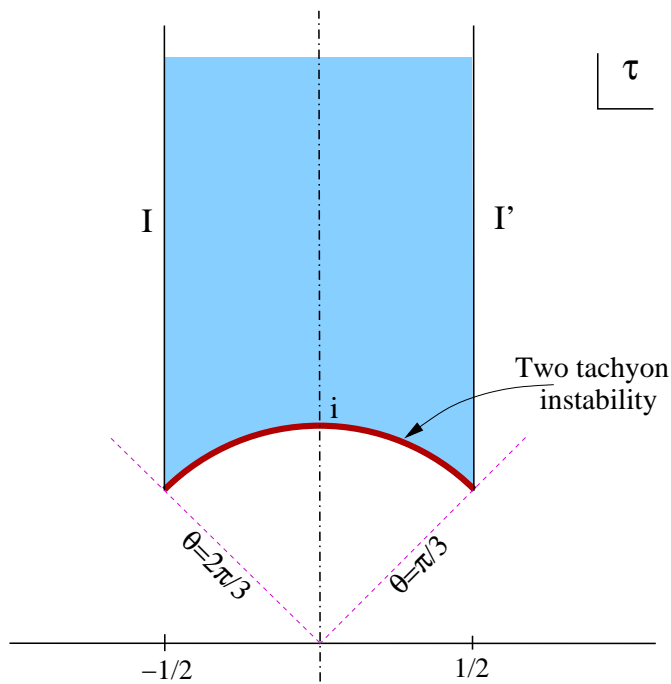}
\caption[]{The shaded area designate the modular domain of ${\bf
    T}^2$ (the sides $I$ and $I'$ are identified).  The two tachyon
  instability develops for torii inhabiting the arc $\tau=\exp(i\,
  \theta)$ with $\pi/3 \le \theta \le 2\pi/3$.}
\label{fig_T2}
\end{figure}
For the example of ${\bf T}^2$ we can find the locus of these
degenerate torii explicitly. They are precisely the torii such
that the basis vectors $\ve_1,\ve_2$ satisfy $|\ve_1|=|\ve_2|$ and
the angle between them, $\theta$, is in the range $\pi/3 \le
\theta \le \pi/2$. In terms of the modular parameter $\tau$ that
means that it lies on the boundary of the modular domain on the
arc $\tau=\exp(i\, \theta)$ with $\pi/3 \le \theta \le 2\pi/3$ see
figure (\ref{fig_T2}).

\section{Indication for critical dimensions with $\Tn$ compactification}
\label{indication}

In the black-string  case there is  a critical dimension
$D^*=13.5$ for the phase transition between the uniform and the
non-uniform states. Namely, for a compactification on $\T1$, the
transition  is of first order  below   $D^*$, while above $D^*$ it
is continuous. In \cite{pertString} it was shown that $D^*$ is
well estimated by $\tD^*$ defined by the intersection
$\mu_{GL}(\tD^*)=\mu_{S}(\tD^*)$. Let us generalize and estimate
such a critical dimension for the $\Tn$ compacification. At the
end of this section we mention other estimates which inquire
whether at given $\mu$ a black-hole can ``fit into'' the compact
dimension (as stressed recently in \cite{MIPark}).

First, considering a single tachyon instability,
 when one of the torus dimensions is much larger than all the
rest, then the situation is exactly analogous to that of the black
string transition. Hence, in this case the critical dimension is
still $13.5$.

Another important example is that of a square $n$-torus where, as
we found in previous section, $n$ tachyons appear simultaneously
below the critical mass. In this case one expects the unstable
modes to deform the horizon along all internal directions. We
would like to know whether the transition to the non-uniform state
is smooth or not. Therefore, generalizing equation (\ref{mu_0})
from the string case,  we define an ``equal area for equal mass''
estimator
\be \label{mu_eqS} \mu_{\rm  S} ={1 \over 16
\pi}\left[{\Omega_{d-2}^{~D-3} \over \Omega_{D-2}^{~d-3} }
  {(d-2)^{(D-3)( d-2)} \over  (D-2)^{(D-2 )(d-3)}}\right]^{1/n},
\ee

\begin{figure}[t!]
\centering \noindent
\includegraphics[width=9cm]{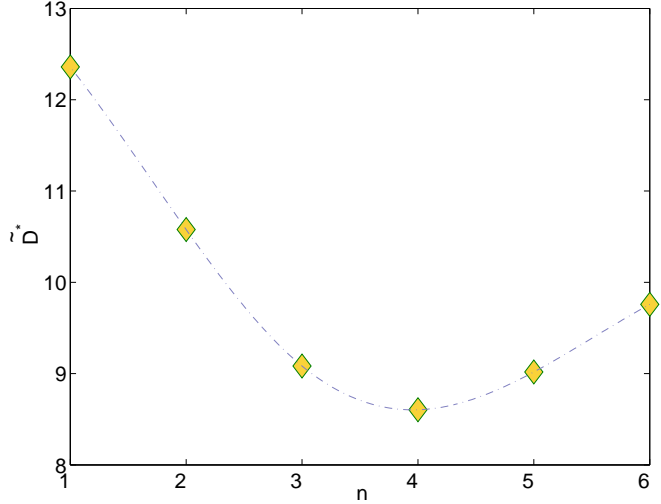}
\caption[]{The dimension $\tD^*$ for which $ \mu_{GL}(\tD^*,n) =
\mu_{\rm S}(\tD^*,n)$ as a function of the dimension $n$  of the
square torus $\Tn$. The closest integer dimension above $\tD^*$
estimates a critical dimension for a change of order in the black
brane phase transition. For $n=1$ this estimate is known to be
short by about 1 from the actual $D^*$: $D^*_{(n=1)} \simeq 13.5
\simeq \tD^*_{(n=1)}+1$. For $3 \le n \le 6$ the estimate is
around $10D$, making it plausible that the actual critical
dimension may be as low as $D^* \leq 11$, where a consistent
theory of quantum gravity is believed to exist.}
\label{fig_2ndOrder}
\end{figure}

The estimate for a critical dimension is obtained by such a
$\tD^*$ that $ \mu_S(\tD^*,n) = \mu_{GL}(\tD^*,n) = \mu_{GL}
(\tD^*-n) $. In figure \ref{fig_2ndOrder} we plot this $\tD^*$ as
a function of the dimension of the internal torus space, $n$.
Above the curve the black hole state is not entropically favorable
at the critical mass. This indicates probably a smooth transition
to a non-uniform brane emerging from the GL-point.
Interestingly, the value of this critical dimension decreases as
more internal dimensions are considered. We learn from figure
\ref{fig_2ndOrder} that for $3\leq n\leq 6$ the total dimension of
the spacetime should be at least $D=10$ to allow a continuous
decay of an unstable black brane to the non-uniform state. This
estimate is somewhat marginal since as we noted above in the $\T1$
case the actual critical dimension is $13.5$ while the estimate is
around $12.5$, namely one less. This indicates that even though
the estimate for $D^*$ is closer to $10$ than in the $\T1$ case,
the actual calculations should be performed in order to establish
whether indeed $D^*=10$ or $11$.

In addition, one should exercise care using the approximate mass
(\ref{mu_eqS}) since it turns out that above certain $D$ and $n$ the
\Sch \ black hole ``does not fit'' into the torus, namely its radius
becomes ``too large'', $2\rho_0/L>1$. In such a case the equation for
$\mu_S$ (\ref{mu_eqS}) is outside its domain of validity. Explicitly,
one obtains
\be \label{ratio} 2\rho_0/L =2
\left({\Omega_{d-2}\over\Omega_{D-2}}\right)^{1\over n} \left({d-2
\over D-2}\right)^{d-2\over n}. \ee
Hence, for the $D=10$ example, the \Sch \ black hole would not fit
into the torus for $n= 5,6$ being \emph{slightly} larger, by $3$ and
$8$ percents respectively, than $L$ (see also \cite{MIPark}).
Actually, there is a little more space in the compact dimension than
appears first due to the ``Archimedes law for caged black holes''
\cite{numericII,GorbonosKol} which tells us that a black hole
``repels'' an amount of space around it that is proportional to its
own size. The effect vanishes in the large $d$ limit, but may cause an
appreciable correction for $5 \le D \lesssim 10$. It would be
interesting to generalize its computation in \cite{GorbonosKol} from
$\T1$ compactification to $\Tn$.

Another interesting estimator can be gotten by considering the \Schw
radius $\rho_0$ for a would be black-hole with mass $\mu_{GL}$. One
finds
 \be
 \rho_0^{~D-3}={G_D \over G_d}\, 
 {(d-2)\Omega_{d-2} \over (D-2)\Omega_{D-2}}\,
 r_0^{~d-3} ~, \ee
 where $r_0$ is the \Schw radius of the critical brane.  For square
 torus $G_D/G_d=L^n$, and taking the large $d$ limit where
 $k_{GL}=r_0(2 \pi/ L) \simeq \sqrt{d}$ one gets \be (D-3) \log\({2\,
   \rho_0 \over L}\) ={D \over 2} \log(D)~ +O(D,d), \ee from which we
 see that for large $D$ (and independently of $d$) the black hole
 cannot fit in the compact dimension at the GL point, and therefore
 the black brane must decay into a different end-state, presumably a
 non-uniform black brane. This estimator can be improved a little by
 replacing the \Schw radius with the radius in conformal coordinates
 $\rho_0 \to \rho_h=\rho_0/2^{2/(d-3)}$ (see \cite{numericI}), and by
 incorporating the ``Archimedes effect'', but these improvements would
 not change the large $D$ result above.

\vspace{0.5cm} \noindent {\bf Acknowledgements}

BK is supported in part by The Israel Science Foundation (grant no
228/02) and by the Binational Science Foundation BSF-2002160. ES
is supported in part by The Israel Science Foundation.

\appendix
\section{Determination of the critical mass $\mu_{GL}(D)$}
\label{appendix_GL}

In this appendix we briefly review the first order calculations of a
Lorentzian perturbation theory around the marginally static GL-string,
$\B_1$, in $D=d+1$ dimensions. In fact, the first order is precisely
what one needs to determine the onset of instability.  An appropriate
perturbation theory was first developed in 5D by Gubser \cite{Gubser},
perfected by Wiseman\cite{GubserPerfected,Wiseman1} in 6D and
generalized in \cite{pertString} to an arbitrary dimension.

The most general ansatz for static black string solutions can be
written as
\bea \label{string_ansatz} ds^2=-e^{2 A} f dt^2 &+&e^{2 B} \left(
f^{-1} dr^2 +dz^2\right) +e^{2 C} r^2 d\Omega_{D-3}^2, \non\
f&=&1-{\left( r_0/r\right)^{D-4}}, \eea
where $A,B$ and $C$ depend on $r,z$ only. When these functions
vanish the metric becomes that of a static uniform black string. We
will work here in units such that  the horizon located at $r_0=1$.

Let us expand the metric functions around the marginally static
uniform solution in powers of some small parameter\footnote{A good candidate for
such a parameter is the dilatonic charge or the tension along the
z-direction\cite{numericI,HO2}, though  for our purposes $\hlm$
even does not have to be specified at this stage.} $\hlm$ such
that in the limit $\hlm \rightarrow 0$ the perturbed string joins
the GL-point
\be \label{expansion}
 A=a(r) \hlm \cos(k z) + O(\hlm^2),~~~  B=b(r) \hlm \cos(k z) + O(\hlm^2),
 ~~~ C=c(r) \hlm \cos(k z) + O(\hlm^2).
 \ee
 The wavenumber $k$ is related to the asymptotic length of the compact
 circle by $k=2\, \pi /L$.

Upon substituting this expansion into the Einstein equations one
obtains a set of ODEs for $a,b$ and $c$. It turns out that in our
case $b$ can obtained algebraically from
\be \label{eq_b} b= \frac{r\, f'\, {a}+2\, \left( D - 3 \right)
\,f\, c +
     2\,r\,f\,\left( a' + \left( D-3 \right) \,c' \right)}{2\,
     \left( D - 3 \right) \,f + r\,f'},
\ee
while for $a$ and $c$ there is an eigenvalue problem to  solve the
set of two coupled linear second order ODEs
\bea \label{eq_ac} -f a''&-& \frac{2\,\left( D-3\right) \,f +
3\,r\,f'
  }{2\,r}a'-\frac{\left( D-3 \right)
  \,f'}{2}c'+k^2 \,a  =0,\\
-f c'' &-& \frac{4\, \left(D-3 \right)\, f\, \left(4 - D +
    \left(D-3\right)\, f \right) - r\, f'\, \left(4\,
    \left(D-3\right)\, f + r\, f' \right) }{r\, \left(2\,
    \left(D-3\right)\, f + r\, f' \right)}\, c' \non &-&\frac{f\,
  \left(2\, \left[2\,(4 - D)+ (D - 3)\, f \right] + r\, f' \right)}
{r\, \left(2\, \left(D-3 \right)\, f + r\, f' \right)} \, a'
+\frac{2\,\left( D-4 \right) \,f'}{r\,\left( 2\,\left( D-3 \right)
\,f
    + r\,f' \right) } \,a\non &-& \frac{2\, \left(D-4\right)\, f'}
{r\, \left(2\, \left(D-3 \right)\, f + r\, f' \right)} \, c + k^2
\, c =0, \eea
subject to the regularity boundary conditions at the horizon, $r=r_0$,
\bea \label{bc_forac} a'&=& {2 \over 3 \,(D-4)}\, k^2 \,a +{D-3
\over 3} \left(-2 \,a +2 \,c - {k^2\over D-4} \,c \right),  \non
 c'&=& 2 \,a - 2 \,c + {k^2\over D-4}\, c,
 \eea
 and at the infinity, $r\rightarrow \infty$, where $a$ and $c$ must
 vanish.

These equations do not have singular points other than horizon and
the spatial infinity unlike the Euclidian perturbation equation
(\ref{chi_eq}). However, there is a price to pay -- instead of a
single second order equation, as in a Euclidean theory, one has to
deal with two coupled equations.

In \cite{pertString} the equations (\ref{eq_ac}-\ref{bc_forac}) were
solved numerically for various $D$'s. The wavenumber $k$ obtained in
this way is precisely that of the marginally tachyonic mode. (Indeed,
the static method that is employed here converges only for this mode.
For $k$ other than $k_{GL}$ there is no perturbed static solution
since above $k_{GL}$ the string is non static being GL-unstable, while
below $k_{GL}$ there is no static perturbed solution either, since
perturbations must decay.)  The calculated values of the critical
wavenumber, $k_{GL}$, are listed in Table \ref{table_kc}.
\begin{table}[h!]
\centering
\noindent
\begin{tabular}{|c||c|c|c|c|c|c|c|c|}\hline
  \rm{D}& 5 & 6&7&8&9&10&11&12 \\ \hline
$k_{GL}$  & .876& 1.27& 1.58& 1.85&2.09& 2.30 &2.50 &2.69
\\ \hline
\hline
  \rm{D}&13&14&15&16&20&30&50&100 \\ \hline
$k_{GL}$  & 2.87 &3.03&3.19 &3.34&3.89 &5.06& 6.72& 9.75
\\ \hline
\end{tabular}
\caption[]{Numerically computed  static
  mode wavenumbers $k_{GL}$ in units of $r_0^{-1}$. }
\label{table_kc}
\end{table}
%


\end{document}